\pdfoutput=1

\documentclass[11pt]{article}

\usepackage[final]{acl}

\usepackage{times}
\usepackage{latexsym}
\usepackage{multirow}

\usepackage[T1]{fontenc}

\usepackage[utf8]{inputenc}

\usepackage{microtype}

\usepackage{inconsolata}

\usepackage{graphicx}

\usepackage{subcaption}
\usepackage{tikz}

\title{FleSpeech: Flexibly Controllable Speech Generation with Various Prompts}

\author{
 \textbf{Hanzhao Li\textsuperscript{1$\ast$}},
 \textbf{Yuke Li\textsuperscript{1}}\thanks{Equal contribution. Random order.},
 \textbf{Xinsheng Wang\textsuperscript{2}},
 \textbf{Jingbin Hu\textsuperscript{1}},
\\
 \textbf{Qicong Xie\textsuperscript{3}},
 \textbf{Shan Yang\textsuperscript{3}},
 \textbf{Lei Xie\textsuperscript{1}}\thanks{Corresponding authors.}
\\
 \textsuperscript{1}Audio, Speech and Language Processing Group (ASLP@NPU), \\ School of Computer Science, Northwestern Polytechnical University, Xi’an, China
 \\
 \textsuperscript{2}Hong Kong University of Science and Technology, Hong Kong, China \\ 
 \textsuperscript{3}Tencent AI Lab, China
\\
 \small{
   \{\href{lihanzhao.mail@gmail.com}{lihanzhao.mail}, \href{yukeli6479@gmail.com}{yukeli6479}, \href{w.xinshawn@gmail.com}{w.xinshawn}, \href{hujingbin553@gmail.com}{hujingbin553}\}@gmail.com, 
 } \\
 \small{
   \href{lxie@nwpu.edu.cn}{lxie@nwpu.edu.cn}, \ 
   \{\href{jerryqcxie@tencent.com}{jerryqcxie}, \href{shaanyang@tencent.com}{shaanyang}\}@tencent.com
 }
}

\begin{document}
\maketitle
\begin{abstract}

Controllable speech generation methods typically rely on single or fixed prompts, hindering creativity and flexibility. These limitations make it difficult to meet specific user needs in certain scenarios, such as adjusting the style while preserving a selected speaker's timbre, or choosing a style and generating a voice that matches a character's visual appearance. To overcome these challenges, we propose \textit{FleSpeech}, a novel multi-stage speech generation framework that allows for more flexible manipulation of speech attributes by integrating various forms of control. FleSpeech employs a multimodal prompt encoder that processes and unifies different text, audio, and visual prompts into a cohesive representation. This approach enhances the adaptability of speech synthesis and supports creative and precise control over the generated speech. Additionally, we develop a data collection pipeline for multimodal datasets to facilitate further research and applications in this field. Comprehensive subjective and objective experiments demonstrate the effectiveness of FleSpeech. Audio samples are available at https://kkksuper.github.io/FleSpeech/ 


\end{abstract}

\section{Introduction}

Speech synthesis plays a pivotal role in content creation and human-computer interaction. With the advancement of powerful generative models, such as large language models~\cite{Valle, TorToiSe, BASETTS, SEEDTTS, ClaM-TTS} and diffusion models~\cite{audiobox, E2TTS, F5TTS}, speech synthesis has experienced rapid progress in recent years~\cite{ControllableSurvey}. Beyond a focus on realism, there is a growing emphasis on \textit{flexible and controllable} speech synthesis~\cite{MM-TTS}, such as the ability to manipulate the style of generated speech based on textual descriptions~\cite{PromptStyle, SALLE, PromptTTS2, UniStyle}.

\begin{figure}[]
  \centering
  \includegraphics[width=0.5 \textwidth]{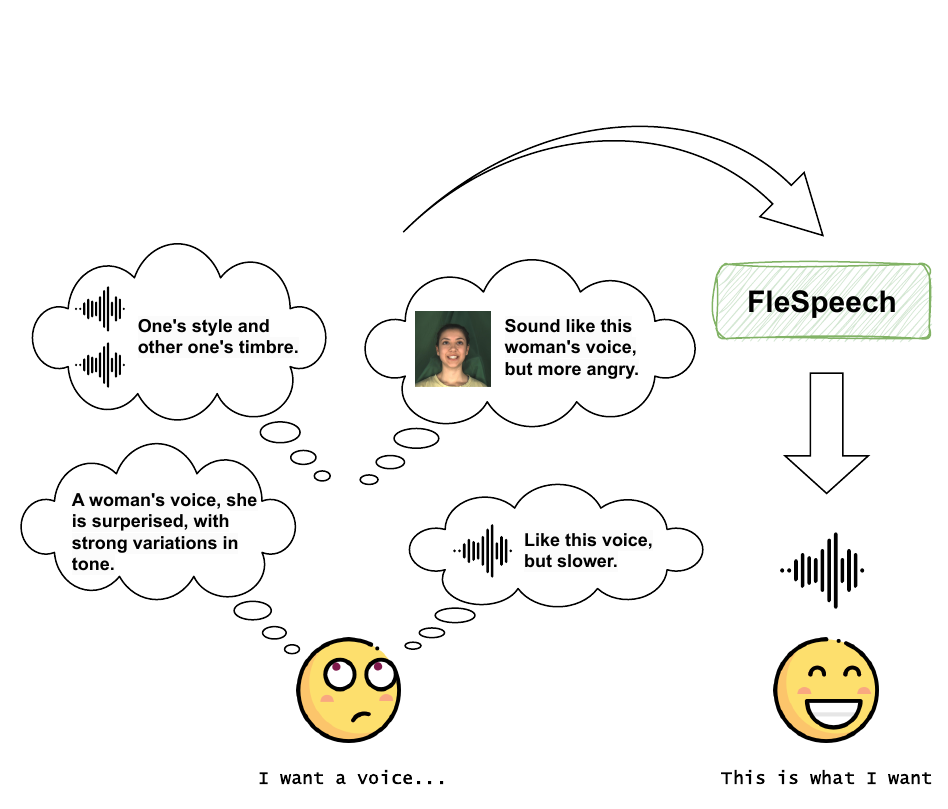}
  \caption {FleSpeech can flexibly generate speech that matches the given prompts.}
  \label{fig:intro}
  \vspace{-10pt}
\end{figure}

Despite the variety of available speech generation control methods, each approach has its inherent limitations. For instance, while speech synthesis based on natural language descriptions offers flexibility, language often struggles to precisely capture all desired attributes, particularly when it comes to describing a speaker's timbre, as textual representations are inherently limited. In contrast, the reference audio-based method can clearly define all attributes but relies on existing audio, which lacks creativity and flexibility. These constraints make it difficult to address specific user needs in certain scenarios, such as adjusting style while preserving a selected speaker timbre or choosing a style and generating a voice that aligns with a character's visual appearance.

To overcome these constraints and move beyond controllable speech synthesis techniques based on a single or a few control methods, we propose a more flexible controllable speech generation method, \textit{FleSpeech}, which supports multiple forms of control and allows for the combination of different control strategies, thereby meeting the flexible control requirements across various scenarios as illustrated in Fig.~\ref{fig:intro}. To this end, we first introduce a multi-stage speech generation framework, with each stage modeling the style and timbre of speech. With this framework, we can provide different prompts at different stages, enabling flexibly controllable speech generation. Second, we propose a multimodal prompt encoder to embed multimodal prompts into a unified representation. Finally, considering the scarcity of multimodal data, we built a data collection pipeline to facilitate research in this area. We will release this data collection pipeline upon the acceptance of this paper.

In summary, the main contributions of this work are as follows:

\begin{itemize}

    \item We propose FleSpeech, a multi-stage speech generation framework that supports multiple prompt inputs to flexibly control different properties of speech. Experiments across different tasks demonstrate both the objective and subjective superiority of this method.

    \item We propose a unified multimodal prompt encoder, which allows us to input any combination of text, audio, and visual modal prompts and operate them in a unified embedding space.

    \item We built a pipeline to facilitate data collection for subsequent multimodal speech generation work.

\end{itemize}

\section{Related Work}

\subsection{Controllable Speech Synthesis}

The employment of category labels, such as speaker identity~\cite{DBLP:conf/interspeech/ChenTRXSZQ20, DBLP:conf/nips/GibianskyADMPPR17} and emotion~\cite{DBLP:journals/corr/abs-1711-05447, DBLP:journals/speech/Lorenzo-TruebaH18}, serves as a prevalent technique for controlling specific speech attributes. To address the limited control capabilities of labels, Skerry-Ryan et al.~\cite{skerry2018towards} introduced a style transfer method based on reference acoustic representation. Subsequently, this reference audio-based approach has gained substantial popularity, particularly in the context of emotion transfer~\cite{li2022cross, lei2022msemotts} and zero-shot TTS~\cite{Valle, ClaM-TTS, du2024cosyvoice}.

To achieve more flexible control, InstructTTS~\cite{yang2024instructtts} and PromptTTS~\cite{guo2023prompttts} are pioneering text description-based speech synthesis, employing natural language to specify the attributes to be controlled. Subsequent efforts~\cite{lyth2024natural, yamauchi2024stylecap, ji2024textrolspeech, PromptTTS2, jin2024speechcraft}  are focused on exploring the use of automated methods to capture more diverse natural language descriptions,  thereby enabling control over an expanded range of attributes.

Additionally, a speaker's facial image can also serve as a form of control information for speech synthesis~\cite{Face2Speech, lee2023imaginary, wang2022anyonenet, FVTTS}. Specifically, AnyoneNet~\cite{wang2022anyonenet} employs face embeddings, projecting them into the same embedding space as reference audio embeddings. This approach aims to generate voices that align with the character's visual appearance, thus facilitating the production of speaker videos that incorporate speech, derived from a single facial image and accompanying text.



Most recently, research has begun to explore control methods beyond single-modality-based methods. MM-TTS~\cite{MM-TTS} pioneers a unified framework that accommodates multimodal prompts from text, audio, or facial modalities. Further advancing this field, StyleFusion TTS~\cite{StyleFusion} introduces a multi-prompt framework that leverages both style descriptions and an audio prompt to simultaneously control audio style and timbre. Unlike StyleFusion TTS, which necessitates simultaneous input of both prompts during inference, our proposed FleSpeech accommodates inputs from any number of arbitrary modalities. This flexibility significantly enhances the adaptability and controllability of speech synthesis.



\begin{figure*}[ht]
    \begin{subfigure}{0.39\textwidth}
        \centering
        \includegraphics[width=1.0\textwidth]{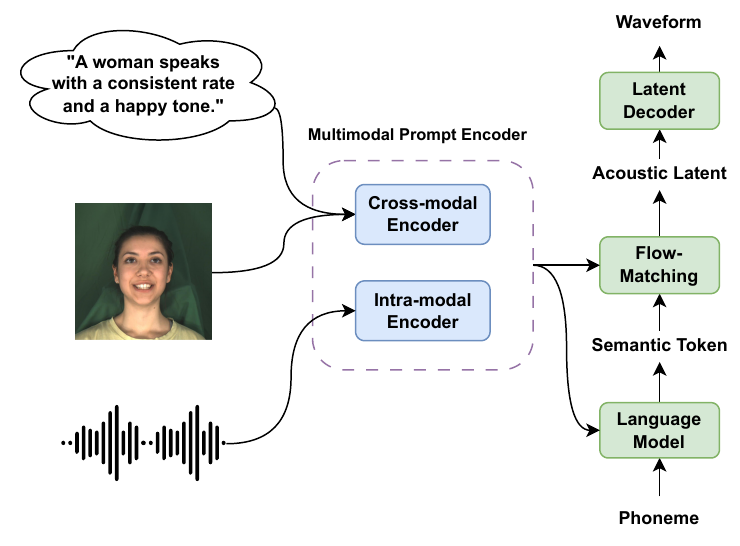}
        \caption{Overall Model Architecture.}
        \label{fig:modela}
    \end{subfigure}%
    \hfill
    \begin{tikzpicture}[overlay]
        \draw[dashed] (0,0.5) -- (0,5); 
    \end{tikzpicture}%
    \hfill
    \begin{subfigure}{0.59\textwidth}
        \centering
        \includegraphics[width=1.0\textwidth]{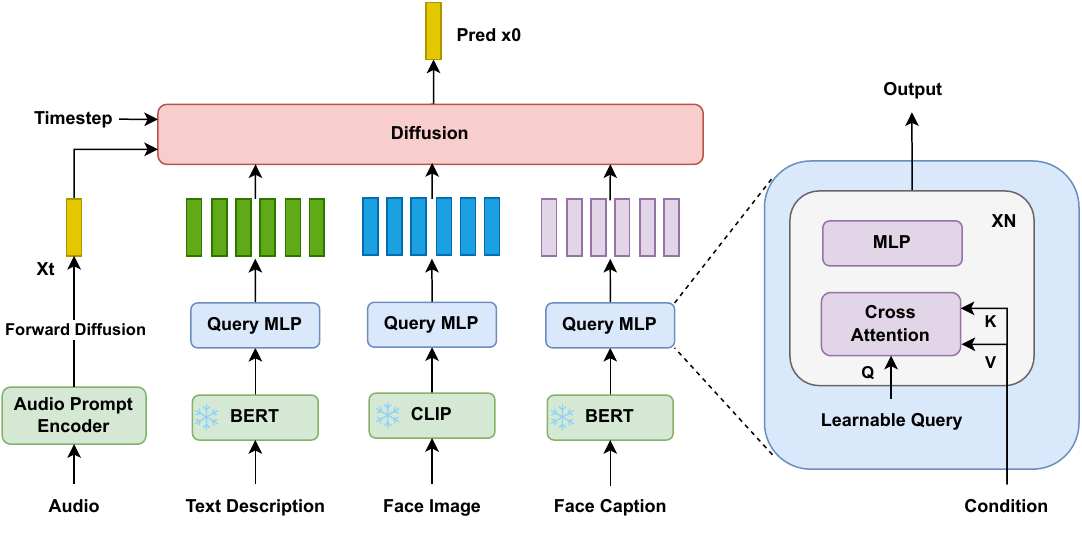}
        \caption{Architecture of Multimodal Prompt Encoder.}
        \label{fig:modelb}
    \end{subfigure}
    \caption{The model architecture of FleSpeech.}
    \label{fig:model}
    \vspace{-10pt}
\end{figure*}

\subsection{Speech Attribute Editing}


Editing speech attributes typically involves modifications to timbre or speaking styles. The former, known as Voice Conversion (VC), specifically aims to transform the timbre to match that of another target speaker while retaining the linguistic information. A typical method employs pre-trained models to extract speaker timbre representations and speech content features, which are then merged to reconstruct the converted speech~\cite{AutoVC, VQMIVC, Expressive-VC}. However, this approach often struggles to generalize to unseen speakers due to model capacity constraints when handling large-scale speech data. To address this challenge, language model-based voice conversion methods have begun to emerge~\cite{wang2024streamvoice, wang2024streamvoice+}.


Instead of changing timbre, style editing focuses on modifying the speech style while preserving linguistic content and timbre. VoxEditor~\cite{VoxEditor} introduces a voice attribute editing model that facilitates the modification of speech style attributes using a given source audio and textual description. Similarly, AudioBox~\cite{audiobox} presents a flow-matching-based framework that enables the restyling of any audio sample through text descriptions. Extending beyond just editing timbre or style, our proposed FleSpeech allows for the simultaneous editing of both speaker timbre and style.


\section{Method}

\subsection{Overview}


FleSpeech is designed to flexibly control the synthesis of speech either through any single-form prompt or a combination of different prompt formats. For instance, it can control style using a text description while managing timbre with reference audio. To facilitate this, as illustrated in Fig.~\ref{fig:modela}, FleSpeech comprises a language model module for semantic token prediction and a Flow Matching-based module for acoustic feature prediction. To handle different forms of prompts, a multimodal prompt encoder (MPE) is proposed. Specifically, MPE is designed to handle prompts in any format, i.e., text, audio, or image, to obtain a unified representation. This unified representation serves as a condition in either the language model or the flow matching module, facilitating targeted control.

Here, we first introduce the language model and flow matching, both of which play crucial roles in speech generation and are classified as components of the multimodal prompt-based speech generator. Subsequently, we describe MPE, which is used to control the generator.

\subsection{Multimodal Prompt-based Speech Generator}



\textbf{Langauge model for semantic generation} 
Inspired by the outstanding performance of language models in speech synthesis tasks~\cite{Valle}, we tokenize speech into semantic tokens and then employ a decoder-only transformer-based language model to predict these tokens. Specifically, the input text is first converted into a phoneme sequence. The language model then takes this phoneme embedding sequence, concatenated with the global condition embedding obtained via MPE, to predict semantic tokens in an autoregressive manner. Details about the model parameters are provided in Appendix ~\ref{append:model_config}.

As for speech tokenization, inspired by Vec-Tok~\cite{VecTok}, our tokenizer employs WavLM~\cite{WavLM}, pre-trained on 94k hour dataset\footnote{https://huggingface.co/microsoft/wavlm-large}, to extract speech features. We then use the K-means clustering method to discretize these features into 300 tokens, primarily associated with linguistic information.



\textbf{Flow matching for acoustic feature generation} 
The absence of acoustic details in semantic tokens results in a gap with the corresponding audio. To bridge this gap, a diffusion transformer based flow-matching-based module, similar to Stable Diffusion 3~\cite{stablediffusion3}, is used to generate acoustic features from semantic tokens, supplemented by the conditional embedding created by the MPE. Details about this module can be found in Appendix ~\ref{append:model_config}.

Compared to pre-designed acoustic features such as the Mel-spectrum, Glow-WaveGAN~\cite{GlowWaveGAN} demonstrates that the acoustic latent representation learned by a variational autoencoder performs better in acoustic feature prediction and vocoder-based speech synthesis processes. Therefore, instead of using the Mel-spectrum as the acoustic feature to be predicted by the flow matching module as in CosyVoice~\cite{du2024cosyvoice}, we adopt WaveGAN implemented in Glow-WaveGAN to extract the latent representation as the acoustic feature via the encoder. The decoder is then used as a vocoder to generate the final audio.




\subsection{Multimodal Prompt Encoder}


The objective of the MPE is to obtain a unified condition embedding based on prompts from multiple modalities. Given that the reference audio contains the most comprehensive information and is always available during the speech generation training process, the core idea behind MPE is to map the representations of textual and visual prompts to the space of reference audio embeddings. To achieve this, following the approach of IP-Adapter~\cite{IPAdapter}, a query-based encoder structure is employed, which uses some learnable query tokens to extract speech-related information from the representations of different prompts. Additionally, due to the many-to-one relationship between reference audio and other prompt modalities, such as multiple voices that correspond to the textual style description "a male speaking loudly and very fast", a diffusion-based method is adopted to model this diversity. Specifically, as shown in the Fig.~\ref{fig:modelb}, the embeddings from different prompt modalities are input into the query-based encoder separately. These embeddings are then concatenated with the noisy audio embedding $x_t$ and fed into the diffusion process. The diffusion model subsequently predicts the ground truth audio embedding $x_0$ through denoising.

\textbf{The reference audio prompt embedding}, serving as the anchor for prompt embeddings from different modalities, captures all time-invariant information, such as style and timbre. Consequently, the embedding created by the reference audio encoder can be directly used as the conditional embedding in speech generation. Similar to Meta-StyleSpeech~\cite{min2021meta}, the reference audio encoder consists of six attention blocks, and the output of the last block is average-pooled to obtain a global audio embedding.

\textbf{The textual prompt embedding} can be derived from either the description of the speaking style or facial visual information. In this case, the description text is embedded using a pre-trained BERT~\cite{BERT}\footnote{https://huggingface.co/google-bert/bert-large-uncased}, which is to capture the semantic information of the descriptions.

\textbf{The visual prompt embedding}, specifically referring to the embedding of face information, is inspired by ID-Animator~\cite{IDAnimator} and aims to capture both static and dynamic information naturally present in face videos. Static information encompasses the facial features of the speaker in a specific frame, such as gender, age, hair colour, and body type, and is closely related to the acoustic features of the speaker. In contrast, dynamic information reflects the speaker's state and behaviour, such as laughing or chatting. This dynamic information complements the static facial features and helps capture nuances that go beyond the capabilities of static images.

MPE is designed to accept inputs from any modality during both training and inference. Embeddings from non-input modalities are masked prior to the diffusion process. Furthermore, given that different speech attributes are modelled at various stages, the parameters of MPE corresponding to token prediction and acoustic feature generation are not shared.


\subsection{Training Strategy}

To address the scarcity of multimodal data, we propose a three-stage training strategy. We use two types of data: 50,000 hours of large-scale low-expressivity speech data from LibriHeavy and 616 hours of high-expressivity speech data collected from the open-source dataset. 

In the first stage, the model is trained on a combination of two datasets to achieve basic speech synthesis capabilities with the large-scale corpus ensuring stability. In the second stage, the model is fine-tuned on high-expressive data to achieve domain alignment. In the third stage, we freeze the generation model backbone and start training the multimodal encoder to enable the model to support modal inputs other than speech prompts. Notably, during this stage,  the multimodal prompt encoder is updated with the generation loss in addition to the diffusion loss. The details of the training objective can be found in Appendix~\ref{append:loss_function}.

\section{Multimodal Dataset}




Due to the scarcity of multimodal controllable speech synthesis data, we propose a method for constructing such a database. 
Compared to existing data, the collected data is not only larger in scale but also includes facial modality with richer facial annotation information. Details about the collected data and comparisons with other multimodal speech synthesis datasets can be found in Appendix~\ref{append:dataset}. 


\textbf{The collection of the talking head video dataset } is based on the CelebV-HQ~\cite{CelebVHQ}, GRID~\cite{GRID}, LRS2~\cite{LRS2}, and MEAD~\cite{MEAD} datasets, which primarily feature talking faces with one person speaking most of the time. After web crawling, the videos are segmented according to the timestamps provided in the dataset. To ensure speech quality, we first apply the S-DCCRN~\cite{lv2022s} model to denoise the crawled videos, retaining only those with a Signal-to-Noise Ratio (SNR) test score greater than 0.6 and a DNSMOS~\cite{reddy2022dnsmos} greater than 2.6. Finally, we use Whisper~\cite{radford2023robust}~\footnote{https://huggingface.co/openai/whisper-large-v3} to get the speech transcription and filter out sentences with fewer than three words. Additionally, the face descriptions are also created, and the details are introduced in section~\ref{sc:face prompt}.

\textbf{The collection of the speech dataset} is based on a large-scale, high-quality TTS dataset, TextrolSpeech~\cite{SALLE}, which concludes emotional content and attribute labels such as gender and emotion. Based on this, we re-caption the speaking style according to the distribution of our entire dataset. This re-caption method is detailed in section~\ref{sc: text prompt}


\subsection{Face Description}
\label{sc:face prompt}



Following ID-Animator~\cite{IDAnimator}, we use both static and dynamic face descriptions. First, we crop all face videos based on timestamp and face range coordinates, selecting a random frame as the face image prompt. This image is processed ShareGPT4V~\cite{chen2025sharegpt4v}~\footnote{https://huggingface.co/Lin-Chen/ShareGPT4V-7B} to generate a static description focused on facial attributes (e.g., gender, age, fatness). To capture the speaker timbre, influenced by facial expressions, we extract video clips and use Video-LLava~\cite{lin2023video} to generate dynamic descriptions focused on facial changes and movements during speech. Finally, we combine both descriptions using a large language model (LLM)~\footnote{We use ChatGPT (gpt-3.5-turbo) as the LLM.} to ensure cohesive and high-quality outputs with relevant details and human-like expression. 

\subsection{Speaking Style Description}
\label{sc: text prompt}

To obtain text descriptions of speaking style, we extract gender and emotion labels from the TextrolSpeech and MEAD datasets. For other talking head video datasets, we use a face gender classification model~\cite{serengil2021lightface}~\footnote{https://github.com/serengil/deepface} to extract gender labels. Acoustic attributes, including pitch, speech rate, and Root Mean Square(RMS) of energy are extracted using the signal processing method. Silent frames are filtered by checking for zero pitch values. In addition, we calculate the mean and variance of pitch to measure the pitch and its fluctuation, and the average RMS to measure the volume.

After feature extraction, we analyze their distribution and apply Mean and One Standard Deviation Splitting to divide each attribute into three intervals: "low," "normal," and "high" intervals. We then use a LLM to generate multiple synonymous words or phrases for each attribute category. Using different prompts, we combine these into single sentences to create various speech style descriptions with the same method. This stage enables the simultaneous generation of multiple speech style descriptions with similar meanings. This method has been shown to provide rich and diverse contextual clues to enhance the effectiveness of zero-shot control.

\section{Experiment Setup}

\subsection{Test Dataset}
To comprehensively evaluate the performance and generalization of the proposed model, two groups of datasets are used for testing. One test set is reserved from the collected multimodal data, which includes 20 voice prompts from TextrolSpeech and 20 facial prompts from the talking head video dataset. The other test set is an out-of-domain dataset from the HDTF dataset~\cite{hdtf}, consisting of image and audio prompts that undergo the same data processing procedures as the training set. Additionally, we selected 16 emotional audio and image prompts from the MEAD dataset to evaluate emotion accuracy. The synthesized transcripts were derived from a random selection of 100 sentences from the multimodal dataset.

\subsection{Evaluation Metrics}

\textbf{Objective metrics} includes Word Error Rate (WER), Speaker Similarity (SPK-Sim), UTMOS~\cite{saeki2022utmos}~\footnote{https://github.com/tarepan/SpeechMOS}, Emotion Accuracy, Gender Accuracy, and other speech attribute accuracy. Details about these objective metrics can be found in Appendix~\ref{append: metrics-obj}.

\textbf{Subjective metrics} include the Mean Opinion Score (MOS) to evaluate speech naturalness (N-MOS) and similarity (Sim-MOS). Higher N-MOS means better naturalness while higher Sim-MOS indicates better similarity with the specific target. Details about the subjective metrics can be found in Appendix~\ref{append: metrics-subj}

\section{Experimental Results}

We evaluated FleSpeech using both single-type prompts and various combinations of prompt types. Additionally, the extended capabilities of FleSpeech, including speech editing and voice conversion, were also assessed. The introduction to the various comparison methods, including MM-TTS~\cite{MM-TTS}, Salle~\cite{SALLE}, NaturalSpeech2~\cite{NaturalSpeech2}, and PromptTTS2~\cite{PromptTTS2} can be found in the Appendix~\ref{append:baseline}.

\begin{table*}[!htbp]
\centering
\caption{Experimental results on speech generation based on a single prompt. $^\diamondsuit$ means the results are obtained from the authors. $^\dag$ means the reproduced results.}
\label{tab:singleprompt}
\renewcommand{\arraystretch}{1.2}
\setlength{\tabcolsep}{5pt}
\scalebox{0.6}{
\begin{tabular}{ll|cccccc|ccc|cc}
\hline
\multirow{2}{*}{Prompt} & \multirow{2}{*}{Model} & \multicolumn{6}{c|}{Accuracy(\%)$\uparrow$}                                      & \multirow{2}{*}{WER(\%)$\downarrow$} & \multirow{2}{*}{SPK-Sim$\uparrow$} & \multirow{2}{*}{UTMOS$\uparrow$} & \multirow{2}{*}{N-MOS$\uparrow$} & \multirow{2}{*}{Sim-MOS$\uparrow$} \\
                            &                        & Emotion       & Gender        & Speed         & Pitch         & Fluctuation   & Volum         &                                                   &                                                 &                                               &                                               &                                                 \\ \hline
\multirow{4}{*}{Text}       & MM-TTS$^\diamondsuit$               & 58.3          & -             & -             & -             & -             & -             & 13.2                                              & -                                               & 1.311                                         & 3.25 $\pm$ 0.08                               & 3.32 $\pm$ 0.03                                 \\
                            & SaLLE$^\dag$                  & 22.4          & 55.2          & 58.3          & 53.5          & 56.8          & 61.7          & 27.2                                              & -                                               & 1.764                                         & 3.02 $\pm$ 0.11                               & 3.17 $\pm$ 0.09                                 \\
                            & PromptTTS2$^\dag$             & 63.5          & 82.6          & 94.6          & 90.6          & 83.2          & \textbf{95.2} & 8.7                                               & -                                               & 1.778                                         & 3.91 $\pm$ 0.08                               & 3.61 $\pm$ 0.07                                 \\
                            & FleSpeech              & \textbf{66.7} & \textbf{89.3} & \textbf{95.1} & \textbf{93.3} & \textbf{95.5} & 92.9          & \textbf{7.5}                                      & -                                               & \textbf{2.351}                                & \textbf{3.95 $\pm$ 0.09}                      & \textbf{4.05 $\pm$ 0.07}                        \\ \hline
\multirow{3}{*}{Audio}      & MM-TTS$^\diamondsuit$                 & 58.8          & 79.3          & -             & -             & -             & -             & 12.8                                              & 0.553                                           & 1.430                                         & 3.56 $\pm$ 0.12                               & 3.38 $\pm$ 0.10                                 \\
                            & NaturalSpeech2$^\dag$         & 64.4          & 88.1          & -             & -             & -             & -             & 7.6                                               & 0.663                                           & 2.602                                         & 3.84 $\pm$ 0.04                               & 3.52 $\pm$ 0.04                                 \\
                            & FleSpeech              & \textbf{66.8} & \textbf{89.9} & -             & -             & -             & -             & \textbf{5.8}                                      & \textbf{0.725}                                  & \textbf{2.835}                                & \textbf{3.94 $\pm$ 0.04}                      & \textbf{3.75 $\pm$ 0.06}                        \\ \hline
\multirow{6}{*}{Face}       & MM-TTS$^\diamondsuit$                 & 56.6          & 70.6          & -             & -             & -             & -             & 17.2                                              & 0.572                                           & 2.155                                         & 3.01 $\pm$ 0.04                               & 3.08 $\pm$ 0.09                                 \\
                            & PromptTTS2$^\dag$             & 63.2          & 72.7          & -             & -             & -             & -             & 11.1                                              & \textbf{0.643}                                  & \textbf{2.643}                                & 3.73 $\pm$ 0.08                               & 3.88 $\pm$ 0.05                                 \\
                            & FleSpeech              & \textbf{64.5} & \textbf{87.3} & -             & -             & -             & -             & \textbf{7.0}                                      & 0.629                                           & 2.457                                         & \textbf{3.91 $\pm$ 0.08}                      & \textbf{3.96 $\pm$ 0.07}                        \\
                            & w/o Face dyn-cap       & 64.0          & 87.1          & -             & -             & -             & -             & 7.1                                               & 0.629                                           & 2.393                                         & 3.82 $\pm$ 0.06                               & 3.91 $\pm$ 0.03                                 \\
                            & w/o Face cap           & 63.0          & 83.7          & -             & -             & -             & -             & 7.2                                               & 0.631                                           & 2.442                                         & 3.72 $\pm$ 0.06                               & 3.83 $\pm$ 0.06                                 \\
                            & w/ FaceNet emb         & 58.5          & 63.8          & -             & -             & -             & -             & 8.2                                               & 0.560                                           & 2.524                                         & 3.58 $\pm$ 0.04                               & 3.25 $\pm$ 0.08                                 \\ \hline
\end{tabular}}
\end{table*}

\begin{table*}[!htbp]
\centering
\caption{Experimental results on speech generation based on multiple prompts.}
\label{tab:multiprompt}
\renewcommand{\arraystretch}{1.2}
\setlength{\tabcolsep}{5pt}
\scalebox{0.65}{
\begin{tabular}{lcc|cccccc|ccc}
\hline
\multirow{2}{*}{Model} & \multirow{2}{*}{Text2Token} & \multirow{2}{*}{Token2Latent} & \multicolumn{6}{c|}{Accuracy(\%)$\uparrow$} & \multirow{2}{*}{WER(\%)$\downarrow$} & \multirow{2}{*}{SPK-Sim$\uparrow$} & \multirow{2}{*}{UTMOS$\uparrow$} \\
                       &                             &                               & Emotion  & Gender  & Speed & Pitch & Fluctuation & Volum &                                                   &                                                 &                                               \\ \hline
FleSpeech              & Text                        & Audio                         & 66.1     & 85.4    & 95.8  & 92.0  & 95.3        & 94.0  & 7.0                                               & 0.706                                           & 2.557                                         \\
FleSpeech              & Text                        & Face                          & 64.9     & 86.3    & 95.2  & 93.7  & 94.9        & 96.4  & 7.2                                               & 0.610                                           & 2.598                                         \\
FleSpeech              & Audio                       & Audio                         & 62.7     & 85.8    & -     & -     & -           & -     & 5.9                                               & 0.702                                           & 3.008                                         \\
FleSpeech              & Audio                       & Face                          & 63.3     & 86.1    & -     & -     & -           & -     & 6.1                                               & 0.603                                           & 2.760                                         \\
w/o Face cap           & Audio                       & Face                          & 63.1     & 81.3    & -     & -     & -           & -     & 6.5                                               & 0.610                                           & 2.667                                         \\ \hline
\end{tabular}}
\vspace{-10pt}
\end{table*}

\subsection{Single-Prompt Controllable TTS}

To evaluate  FleSpeech's single-prompt control capabilities, we compared it with other models using text, face image, or audio as the prompt. We also conducted an ablation study to show the effectiveness of FleSpeech's design.


\subsubsection{Comparsion with Other Methods}

\textbf{Speech generation with text prompt} was conducted using a set of text prompts with various emotional and prosodic attributes. As shown in the \textit{Text} section of Table~\ref{tab:singleprompt}, FleSpeech achieved excellent results in terms of different style attributes and emotional accuracy. Subjective testing results indicate that the speech generated by FleSpeech closely follows the text prompts and exhibits high naturalness. 

\textbf{Speech generation with audio prompt} is presented in the \textit{Audio} section of Table~\ref{tab:singleprompt}. Compared to MM-TTS, FleSpeech demonstrates significantly better speaker similarity, primarily due to the model capacity of the large-scale speech synthesis system. Furthermore, FleSpeech outperforms NaturalSpeech2 in terms of emotion accuracy, gender accuracy, and speaker similarity, highlighting that its multi-stage framework is more effective at capturing various attributes, such as style and tone from the audio prompts. With the cascading structure of LM and flow matching, FleSpeech has significantly improved naturalness and audio quality.


\textbf{Speech generation with face prompt} presented in the \textit{Face} section of Table~\ref{tab:singleprompt} showcases that FleSpeech achieved optimal performance across most metrics except for speaker similarity. This is primarily due to the absence of an explicit objective relationship between speaker timbre and facial features. Instead, the matching is more subjective in nature. Subjective results indicate that the speech generated by FleSpeech has a higher correlation with the facial images, suggesting its ability to capture key information from the face and synthesize matching speaker timbre.


\subsubsection{Ablation Study}


To evaluate the effectiveness of face captions, an ablation study was conducted, which can be found in the \textit{Face} section of Table~\ref{tab:singleprompt}. We first removed the dynamic attributes of the face description (w/o Face dyn-cap), which resulted in a sharp decline in emotional similarity, indicating a reduced ability of the model to capture emotional information from the face. Moreover, when we eliminated both the static and dynamic attributes of the face description (w/o Face cap), the model relied solely on Clip representations for speaker-timbre-related information. The experimental results show a comprehensive decline in terms of all metrics, demonstrating the effectiveness of combining Clip and facial descriptions. Finally, we replaced Clip with FaceNet (w/ FaceNet emb), a facial recognition model capable of extracting embeddings that represent unique attributes among different individuals for face-driven speech synthesis. The experimental results indicated that FaceNet’s ability to capture facial information is insufficient for synthesizing speech corresponding to the face prompt.

\begin{figure}[]
  \centering
  \includegraphics[width=0.45 \textwidth]{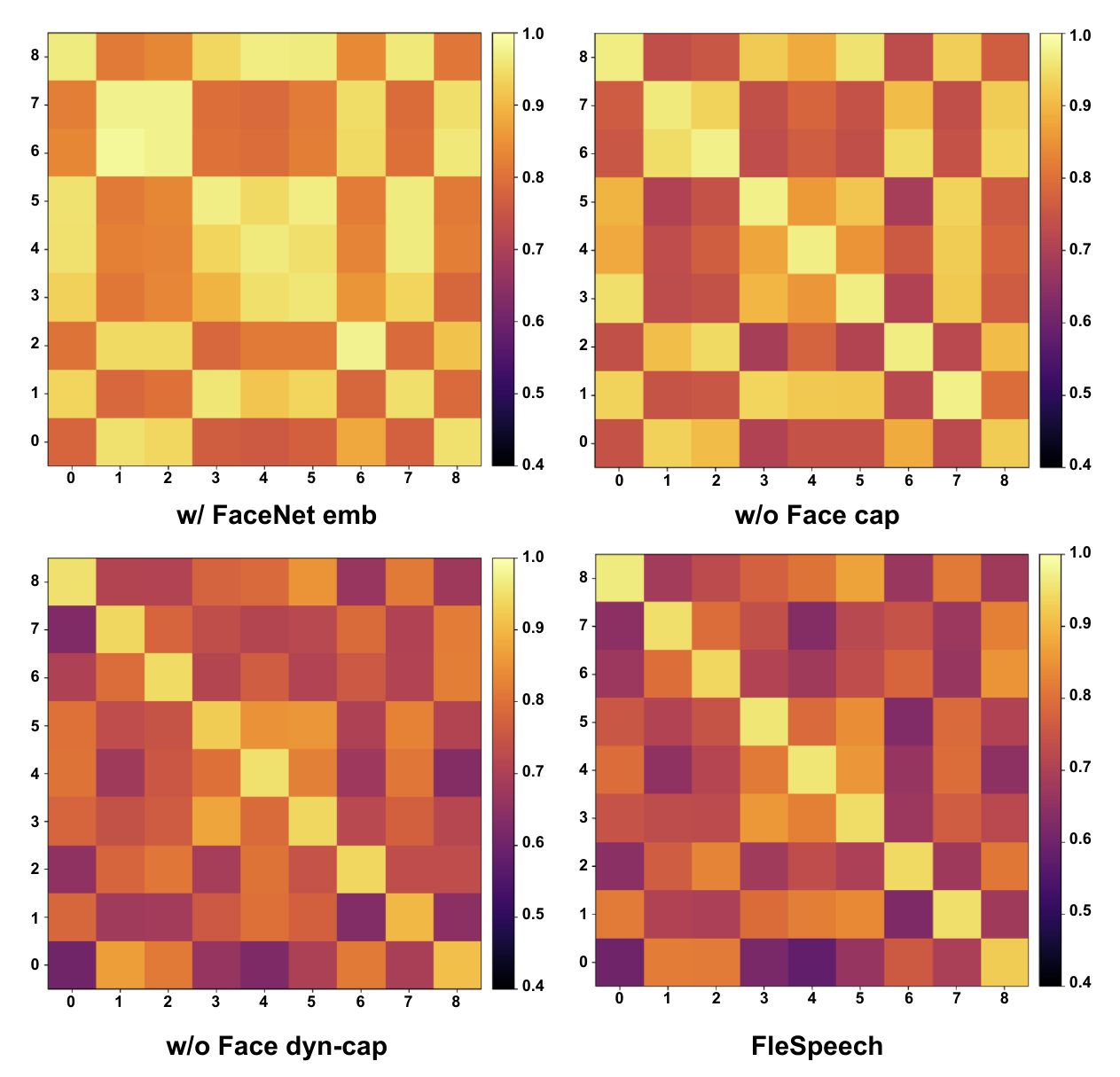}
  \caption {Cosine similarity matrix of speaker embeddings between face-prompt-based synthesized speech and ground-truth speech. The horizontal axis represents different synthesized speech, while the vertical axis represents ground-truth speech. The diagonal indicates that the image prompt and ground-truth speech are from the same speaker. Lighter colors indicate higher similarity.}
  \label{fig:spkemb}
  \vspace{-15pt}
\end{figure}

We further visualized the speaker embedding similarity matrix between different generated sentences. As shown in Fig.~\ref{fig:spkemb}, compared to the results with w/ FaceNet emb, Clip (i.e., w/o Face cap) exhibits higher speaker consistency, indicating the effectiveness of the Clip encoder in extracting implicit representations. By gradually adding static or dynamic face captions, the colors outside the diagonal gradually deepen, indicating a stronger binding between facial images and speaker timbre. FleSpeech demonstrates the highest speaker consistency, highlighting the effectiveness of combining Clip with dynamic and static captions.


\subsubsection{Overall Analysis}
 
In addition to individual tasks, we conducted an overall analysis of the different experimental results. The comparative results in various sections of table~\ref{tab:singleprompt} indicate that the audio modality achieves the highest accuracy in terms of emotion and gender, followed by text. This suggests that audio provides the most fine-grained information, and through the text prompt encoder, the model can effectively extract relevant speech attributes from textual descriptions. Image prompts, on the other hand, are generally less discernible, leading to a decrease in accuracy. Moreover, the WER and UTMOS of speech generated from text prompts show a significant decline, which may be attributed to the one-to-many problem, especially in the text modality, where a larger sample space results in poorer stability. Finally, despite being trained on a small-scale dataset, we observed that MM-TTS using face prompt outperforms the audio prompt in terms of SPK-Simi and UTMOS. This reflects the generalization advantage of the face prompt, considering the complex acoustic environments present in the audio prompt.

\subsection{Multi-Prompt Controllable TTS}


To evaluate the unique flexible control capability of  FleSpeech, we assessed its performance using multiple prompts. Specifically, we provide different prompts at various stages to control different speech aspects. We examined four combinations of prompts. To validate the effectiveness of each stage, we included emotional or neutral prompts in the first stage and only neutral prompts in the second stage. As shown in Table~\ref{tab:multiprompt}, compared to using a single prompt for control, FleSpeech effectively controls style and emotional attributes while reproducing the timbre of the target speaker despite some performance loss.

Additionally, we removed the facial caption(w/o Face cap) in the combination of audio and prompts. We observed a significant decrease in gender accuracy, which indicates that fine-grained information provided by the audio prompt affects speaker timbre modeling in the second stage. The experimental results demonstrate that incorporating the face caption can alleviate the impact of audio prompts, leading to higher consistency with the face prompt.

Furthermore, by comparing the results of different tasks, we found that the WER and UTMOS are highest for the model using two audio prompts, while models using text as the first-stage prompt have the lowest values. This further indicates a negative correlation between the diversity and stability of the speech attribute space. Moreover, models using text as the first-stage prompt generally achieve higher SPK-Sim compared to those using audio modality. This suggests that more fine-grained information in the first stage can influence the speaker timbre modeling in the second stage.

\subsection{Extensibility}

In addition to speech synthesis, we conducted additional experiments on other tasks to evaluate the scalability of FleSpeech.

\subsubsection{Speaking Style Editing}


Speaking style editing refers to modifying speech attributes without altering the content or speaker timbre. To edit the attribute of a given utterance based on the text description, the transcription of this utterance obtained via Whisper and the text description can be used as the input for the language model. Then this utterance can work as the audio prompt for the second stage.
We compared our method with Audiobox~\cite{audiobox}, a unified audio generation model based on flow matching that can redesign the provided audio examples using natural language instructions. As shown in Table~\ref{tab:editing}, FleSpeech achieves satisfactory results. Regarding emotional expression, FleSpeech scores lower, primarily because Audiobox incorporates non-verbal sounds, such as laughter, which enhance emotional perception.

\begin{table}[!htbp]
\vspace{-5pt}
\centering
\caption{Experimental results in speaking style editing.}
\vspace{-4pt}
\label{tab:editing}
\renewcommand{\arraystretch}{1.2}
\setlength{\tabcolsep}{2pt}
\scalebox{0.62}{
\begin{tabular}{l|ccccc|cc}
\hline
\multirow{2}{*}{Model} & \multicolumn{5}{c|}{Accuracy(\%)$\uparrow$}                      & \multirow{2}{*}{WER(\%)$\downarrow$} & \multirow{2}{*}{SPK-Sim$\uparrow$} \\
                       & Emotion       & Speed         & Pitch         & Fluctuation   & Volum         &                                                   &                                                 \\ \hline
Audiobox               & \textbf{66.3} & 83.3          & \textbf{98.3} & 83.3          & 83.3          & 8.4                                               & 0.712                                           \\
FleSpeech              & 63.6          & \textbf{91.6} & \textbf{98.3} & \textbf{91.6} & \textbf{91.6} & \textbf{7.2}                                      & \textbf{0.745}                                  \\ \hline
\end{tabular}
}
\vspace{-15pt}
\end{table}

\subsubsection{Voice Conversion}

FleSpeech allows for the speaker timbre editing by facial caption when given a facial image and its corresponding caption. For instance, it can explicitly specify attributes such as the speaker's age, race, and fatness, which have been previously proved to be associated with speaker timbre~\cite{ stathopoulos2011changes, souza2018body, yang2022speaker}. We evaluate the effectiveness of these edits through accuracy testing and subjective preference assessments. The MOS indicates the degree of match, with higher scores reflecting better alignment. Preference indicates perceived accuracy, where participants choose which audio, before or after editing, better matches the edited facial caption. The test results are shown in Table~\ref{tab:vc}, where FleSpeech achieves an editing accuracy exceeding 70\%, demonstrating its capability to effectively edit speaker-timbre-related attributes to match facial features. The subjective scores further corroborate this conclusion. Additionally, the accuracy for age is higher than for BMI, suggesting that age is more perceptible in facial images.

\begin{table}[!htbp]
\vspace{-5pt}
\centering
\caption{Experimental results in voice conversion.}
\label{tab:vc}
\vspace{-4pt}
\setlength{\tabcolsep}{2pt}
\renewcommand{\arraystretch}{1.2}
\scalebox{0.75}{
\begin{tabular}{lccc}
\hline
Characteristic & \multicolumn{1}{l}{Acc(\%)$\uparrow$} & \multicolumn{1}{c}{MOS$\uparrow$} & \multicolumn{1}{l}{Preference(\%)$\uparrow$} \\ \hline
BMI        & 72.6                    & 3.75 ± 0.04            & 62.4                            \\
Age            & 81.0                     & 3.87 ± 0.08                      & 74.1                            \\
Race           & 75.3                     & 3.83 ± 0.06                     & 66.5                            \\ \hline
\end{tabular}
}
\vspace{-12pt}
\end{table}

\section{Conclusion}

In this work, we propose a flexible and controllable speech generation framework called FleSpeech. Specifically, we implement a two-stage speech generation framework composed of a language model and a flow matching module, allowing for flexible control by providing different prompts at various stages. Additionally, we introduce a multimodal prompt encoder that can accept prompts from different modalities and embed them into a unified style space, enabling more adaptable prompting. Comprehensive subjective and objective experiments demonstrate the effectiveness of FleSpeech.

\section*{Limitation}
Although our approach successfully achieves flexible control over speech attributes, it is important to acknowledge its limitations. First, the information extracted from face images is limited. Many unexplored aspects, such as accent, are related to speaking style and restrict the matching accuracy between face and speech. Second, the relatively small scale of our collected dataset limits the control over additional attributes, such as background sound. Despite these limitations, our FleSpeech has taken an important step toward a more flexible and controllable speech generation system.


\bibliography{custom}

\appendix

\section{Model Configurations}
\label{append:model_config}

The language model for semantic prediction adopts the LLaMA architecture with 16 layers and 16 attention heads. The hidden size and intermediate size are 1024 and 4096, respectively. The flow matching for the acoustic feature prediction is based on the DiT architecture with 12 layers, 12 attention heads, and a hidden dimension of 768. For MPE, the number of queries in QueryMLP is set to 16, with 6 layers, 6 attention heads, and an intermediate size of 256. The reference audio encoder consists of 6 attention blocks with a hidden size of 512. During both the training and inference stages, the length of the audio prompt is fixed at 6 seconds. 

The diffusion model consists of a diffusion process and a denoising process. For the diffusion process, given the audio embedding $x_0$, the forward diffusion process transforms it into Gaussian noise under the noise schedule $\beta$ as follows:
\begin{equation} 
\mathrm{d}x_t=-\frac{1}{2}\beta_{t}x_{t}\mathrm{t}+\sqrt{\beta_{t}}\mathrm{d}\omega_{t}, t\in [0,1]
\end{equation}
For the denoising process, the denoising process aims to transform the noisy representation $x_t$ to the audio embedding $x_0$ by the following formulation~\cite{DBLP:conf/iclr/0011SKKEP21}.
\begin{equation} 
\mathrm{d} x_{t}=-\frac{1}{2}\left(x_{t}+\nabla \log p_{t}\left(x_{t}\right)\right) \beta_{t} \mathrm{~d} t, \quad t \in[0,1]
\end{equation}
The diffusion module is trained to estimate the gradients of log-density of noisy data ($\bigtriangledown log p_{t}(z_{t}) $) by predicting the origin audio embedding $x_0$, conditioned on the embeddings from different prompt modalities, noised audio embedding, and diffusion step $t$ that indicates the degree of noise in the diffusion model. 

\begin{table*}[!htbp]
\centering
\caption{Comparison between public datasets for controllable speech generation. Rec means recording, You means youtube, Pod means podcast}
\label{tab:append-dataset}
\renewcommand{\arraystretch}{1.5}
\setlength{\tabcolsep}{5pt}
\scalebox{0.72}{
\begin{tabular}{cccccc}
\hline
\textbf{Dataseet} & \textbf{Duration} & \textbf{Clips} & \textbf{Modality}    & \textbf{Audio Source}            & \textbf{Description Form}              \\ \hline
FSNR0             & 26h               & 19k            & Speech               & Internal dataset                 & Style tag                              \\
TextrolSpeech     & 330h              & 236k           & Speech               & Recording, Emotional dataset     & LLM template                           \\
MEAD-TTS          & 36h               & 31k            & Speech, Facial image & Recording                        & LLM template, Face image               \\
Collected data           & 616h              & 449k           & Speech, Facial image & Rec, You, Pod, Emotional dataset & LLM template, Face image, Face caption \\ \hline
\end{tabular}
}
\end{table*}

Both language model and flowing matching module are trained on 8 NVIDIA TESLA V100 GPUs (32GB each) with a batch size of 2 per GPU and a gradient accumulation step of 50. The two models are first trained 600k steps on the LibriHeavy~\cite{Libriheavy} dataset which is a 50,000 hours ASR corpus, followed by an additional 300k steps on a collected multimodal dataset. We optimize the models using the AdamW optimizer, warming up the initial learning rate from $ 1 \times 10^{-7} $ over the first 5k updates to a peak of $3 \times 10^{-4}$, and subsequently applying cosine decay.

\section{Training Objective}
\label{append:loss_function}

In the first stage, the language model performs the next token prediction task and is optimized using the cross-entropy loss. Meanwhile, flow matching reconstructs the hidden layer features and is optimized with \( L_2 \) loss. In the second stage, the MPE is optimized using the \( L_1 \) loss calculated between the output embedding \( Pred \ x_0 \) derived from different prompt modalities and ground-truth embedding \( x_0 \) obtained from the audio modality, in addition to the loss function of the first stage.

To achieve flexible control, the MPE applies masking to the received prompts of different modalities. Specifically, the audio modality prompt, serving as the target for the MPE, remains consistently present. For data containing both text and audio modality prompts, the MPE maps the text prompt embeddings to the audio embeddings without any masking. In the case of data that includes prompts from all three modalities, there is a one-third probability of masking the text style description, a one-third probability of masking the facial image and facial description, and a one-third probability of not applying any masking. This strategy enables the model to accept various combinations of prompts as input during the inference stage.

\section{Details of Collected Data}
\label{append:dataset}

As shown in Table~\ref{tab:append-dataset}, previous work has attempted to construct public datasets for controllable speech generation, but these datasets either have limited size or lack multimodal prompts. In view of this, we constructed a multimodal dataset collection pipeline. Through this pipeline, we collected a multimodal TTS dataset consisting of a 285.9-hour talking head video dataset and a 330-hour speech dataset, totaling approximately 615.9 hours.

\section{Evalution Metrics}

\subsection{Objective Metrics}
\label{append: metrics-obj}

\textbf{WER} is a commonly used metric to assess the intelligibility of generated speech. It is typically calculated by comparing the transcribed text obtained from an Automatic Speech Recognition (ASR) system with the reference text. A lower WER indicates higher intelligibility of the speech. Here, the WER is calculated based on the Whisper~\cite{radford2023robust}~\footnote{https://huggingface.co/openai/whisper-large-v3} model. 

\textbf{SPK-Sim} is used to evaluate the similarity between the generated audio and the reference audio in terms of speaker characteristics. A higher SPK-Sim value indicates greater similarity between the synthesized speech and the reference audio in terms of the speaker's identity. Here, we use WavLM-large~\cite{chen2022large}fine-tuned on the speaker verification task, to obtain speaker embeddings. These embeddings are then used to calculate the cosine similarity between the speech samples of each test utterance and the reference clips. 

\textbf{Emotion Accuracy} is used to measure the model's ability to control emotions. A higher emotion accuracy indicates a stronger ability of the model to control emotions. Here,  emotion2vec+seed~\cite{ma2023emotion2vec}~\footnote{https://huggingface.co/emotion2vec/emotion2vec\_plus\_seed} is adopted to predict the emotion of the synthesized audio and compare it with the given emotion type. 

\textbf{Gender Accuracy} is used to measure the model's ability to control gender. A higher gender accuracy means a better gender control ability. Here, an internal ECAPA-TDNN~\cite{desplanques2020ecapa} model fine-tuned on the gender classification task is adopted. 

For the accuracy of other speech attributes, we utilize the previously mentioned pipeline for style label annotation to extract attribute values and compare their relative magnitudes across different labels. For example, the speech rate associated with the “fast speaking rate” label exceeds that of the “slow speaking rate” label. For face attribute evaluation, we extract speaker embeddings from MPE and use a face classifier~\footnote{https://github.com/lsimmons2/bmi-project} to predict Body Mass Index (BMI). Additionally, we apply the DeepFace~\cite{serengil2021lightface} model to determine gender, race, and age. We then train an MLP-based predictor to infer facial attributes from the speaker embeddings, comparing the predicted attributes against the provided facial descriptions to compute the accuracy.

\subsection{Subjective Metrics}
\label{append: metrics-subj}

In the subjective evaluation, each sample was rated on a scale from 1 to 5, with increments of 0.5 based on its similarity to the reference utterance, where a score of 1 indicates “very bad” and a score of 5 signifies “excellent.” Both Normalized Mean Opinion Score (N-MOS) and Similarity Mean Opinion Score (Sim-MOS) are reported with a 95\% confidence interval. We selected 50 speech samples for each test, which were listened to by 20 listeners for subjective evaluations.

To clarify, Sim-MOS here varied across different tasks, focusing on aspects such as speech style matching with a text prompt, speaker similarity with an audio prompt, and voice-face matching with a facial prompt.

\section{Comparison models}
\label{append:baseline}

To evaluate the performance of FleSpeech, we implemented the following system.

\begin{itemize}

\item \textbf{MM-TTS} \cite{MM-TTS}: A FastSpeech2-based multimodal controllable speech synthesis framework that integrates multimodal inputs into a unified representation space. It supports text descriptions, face images, or speech as prompts. Note that text descriptions in this model are limited to describing the speaker's emotions.

\item \textbf{Salle} \cite{SALLE}: A VALL-E-based text-prompt-driven controllable speech synthesis framework, where text descriptions are concatenated with synthesized phonemes as style prompts.

\item \textbf{NaturalSpeech2} \cite{NaturalSpeech2}: A TTS system with latent diffusion models to enable zero-shot speech synthesis. 

\item \textbf{PromptTTS2} \cite{PromptTTS2}: A NaturalSpeech2-based speech synthesis framework capable of generating speech that aligns with text style descriptions. We extend its function to support the face prompt just as described in the PromptTTS2 appendix. The CLIP model extracts embedding from the face image, which is then fed into the TTS model.

\item \textbf{FleSpeech}~(proposed): Our proposed framework, which adopts a multi-stage training framework and follows a multi-stage training strategy.

\end{itemize}

\section{Visualizing the Relationship between Facial Attributes and Voice}
\label{append:facevisual}
\begin{figure}[]
    \centering
    \begin{minipage}{0.23\textwidth}
        \centering
        \includegraphics[width=\textwidth]{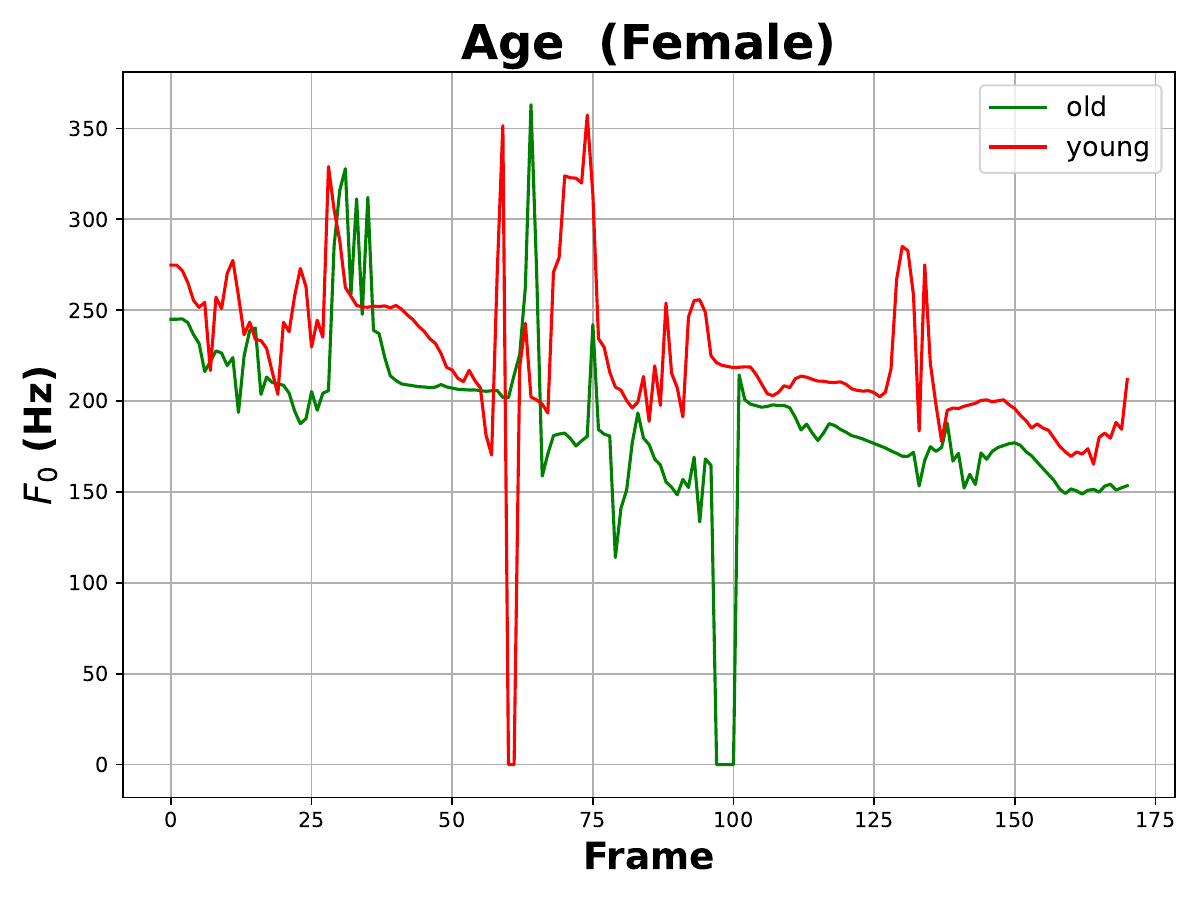} 
    \end{minipage}
    \hfill
    \begin{minipage}{0.23\textwidth}
        \centering
        \includegraphics[width=\textwidth]{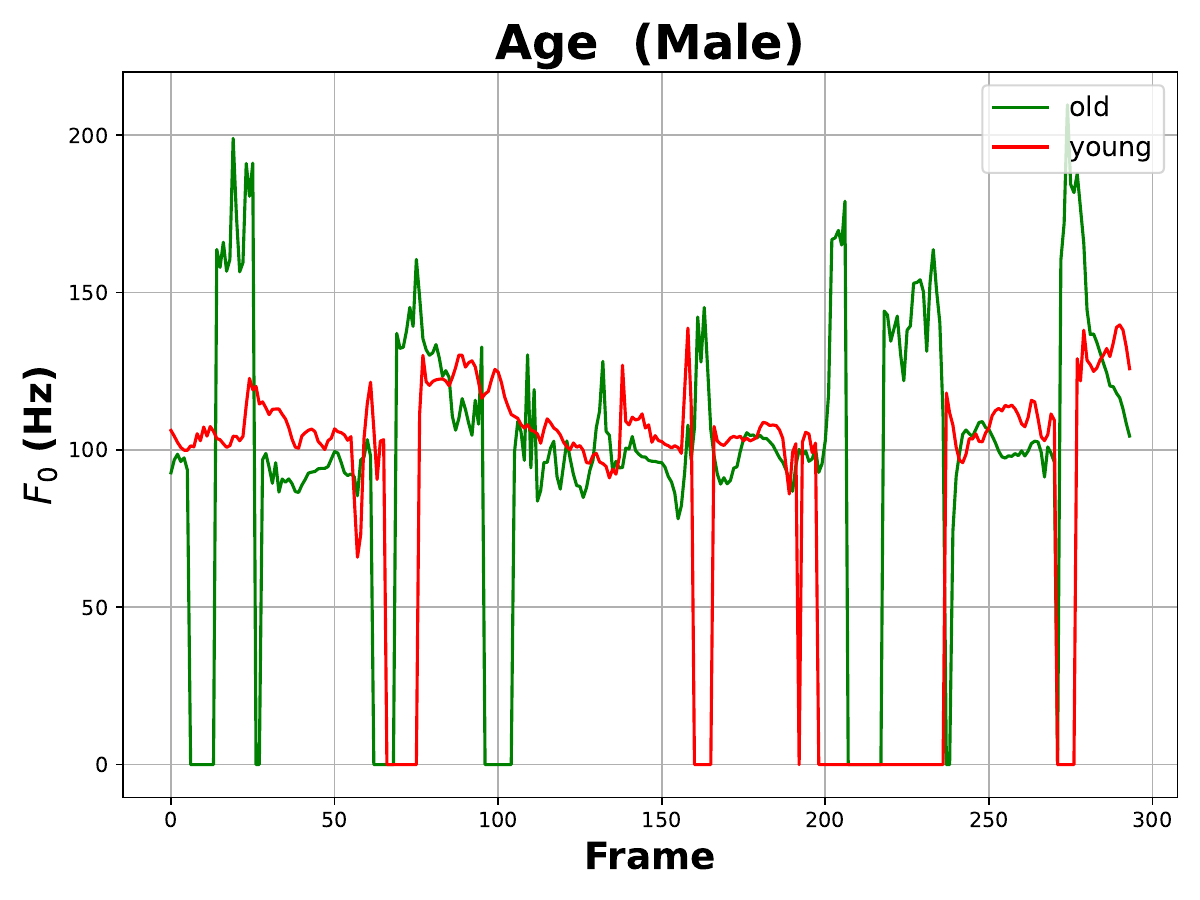} 
    \end{minipage}
    
    \vskip 0.5cm  
    \begin{minipage}{0.23\textwidth}
        \centering
        \includegraphics[width=\textwidth]{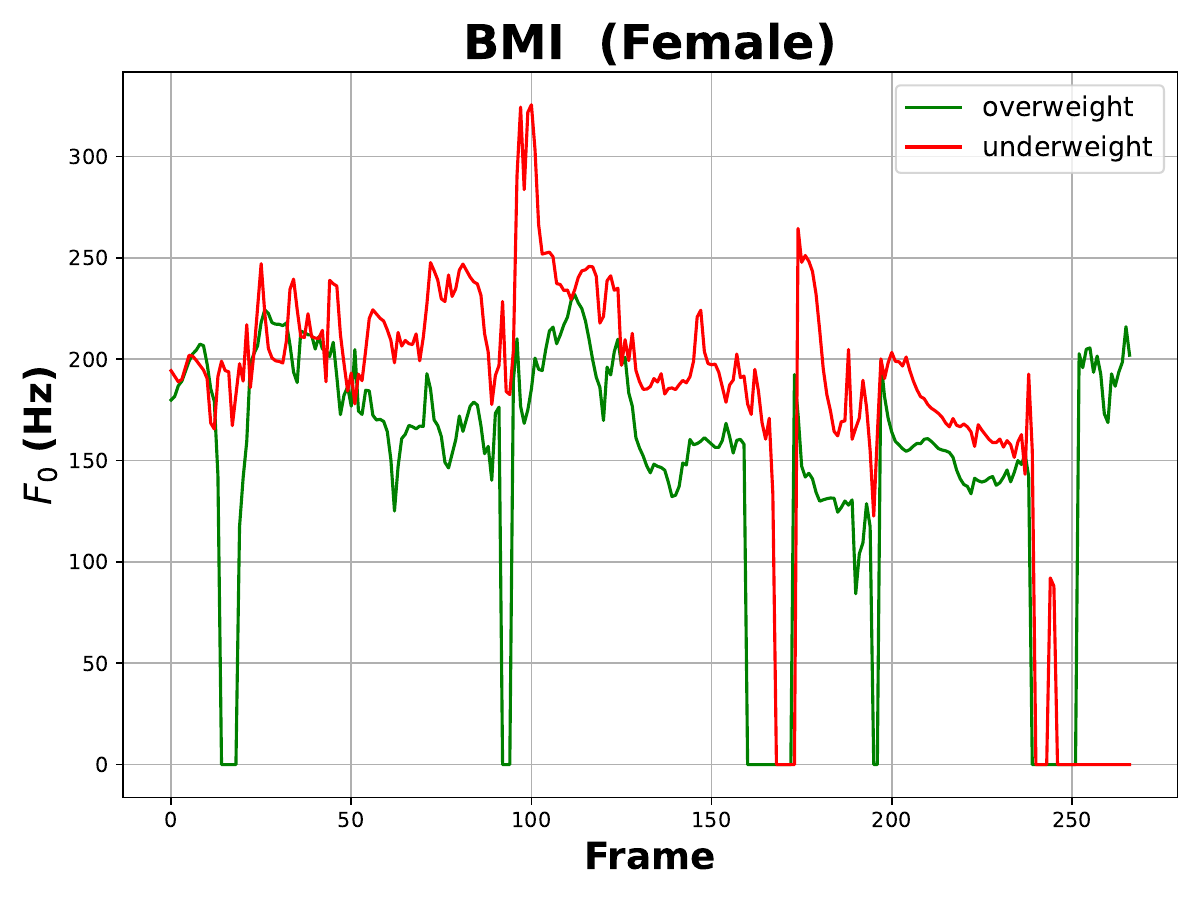} 
    \end{minipage}
    \hfill
    \begin{minipage}{0.23\textwidth}
        \centering
        \includegraphics[width=\textwidth]{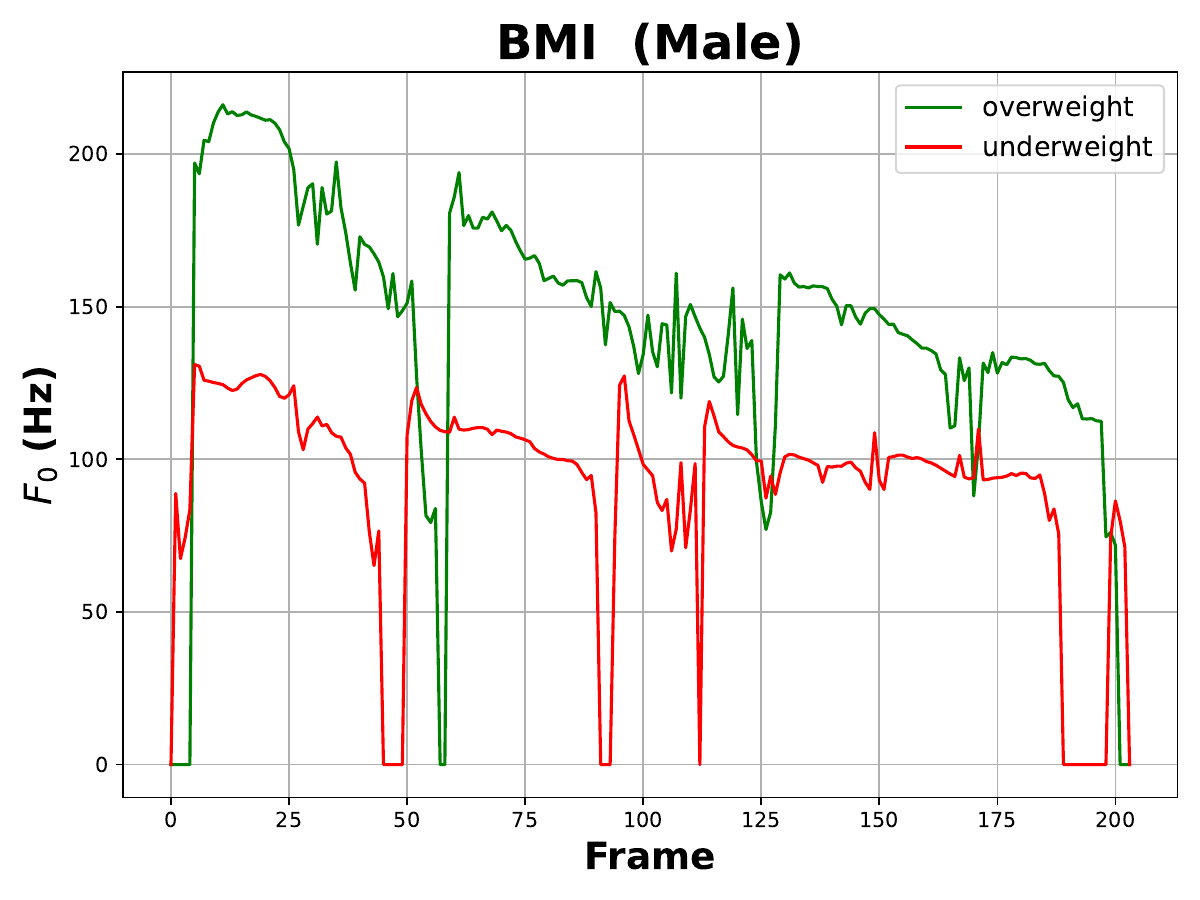} 
    \end{minipage}
    
    \caption{Fundamental Frequency (F0) curve of the speech at different ages and BMI levels groups by gender}
    \label{fig:append-facepitch}
\end{figure}

To further validate that FleSpeech can establish associations between facial attributes and voice characteristics, we extracted the fundamental frequency (F0) from synthesized speech prompted by different BMI and age groups. As shown in the upper two panels of Fig.~\ref{fig:append-facepitch}, the F0 of older women decreases as age increases. In contrast, for elderly males, despite vocal cord atrophy, the F0 tends to increase. The lower two panels of Fig.~\ref{fig:append-facepitch} reveal that being overweight tends to cause articulation difficulties, leading to a decrease in F0 for females, whereas males experience an increase in F0. These findings align with conclusions from prior research, indicating that the proposed FleSpeech can capture variations in speech characteristics across different ages and BMI levels.

\section{Visualization of MPE Embedding Space}
\label{append:mpevisual}

We design MPE to encode prompts from different modalities into a unified space. To validate this, we utilized the MPE to extract embeddings corresponding to each single-modality prompt. Considering that the MPEs in the language model and flow matching do not share parameters, we conducted analyses on both. The test set comprised 2000 randomly selected sentences containing prompts from all three modalities, including 20 speakers with 200 sentences each. The MPE outputs are projected to 2D by t-SNE~\cite{van2008visualizing}. Each color represents a modality.

\begin{figure}[]
    \centering
    \begin{minipage}{0.23\textwidth}
        \centering
        \includegraphics[width=\textwidth]{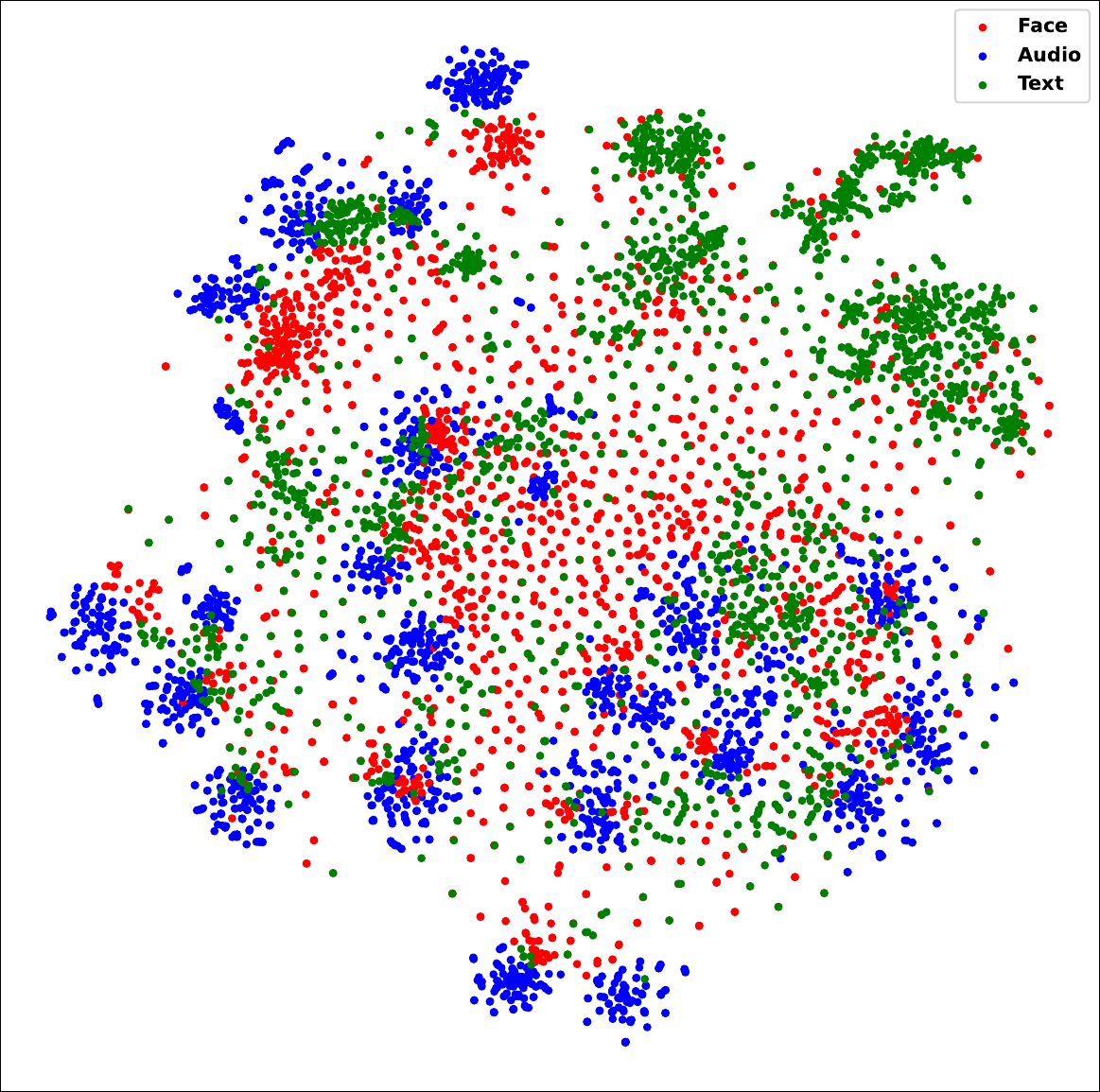} 
        \subcaption{MPE in language model}
    \end{minipage}
    \hfill
    \begin{minipage}{0.23\textwidth}
        \centering
        \includegraphics[width=\textwidth]{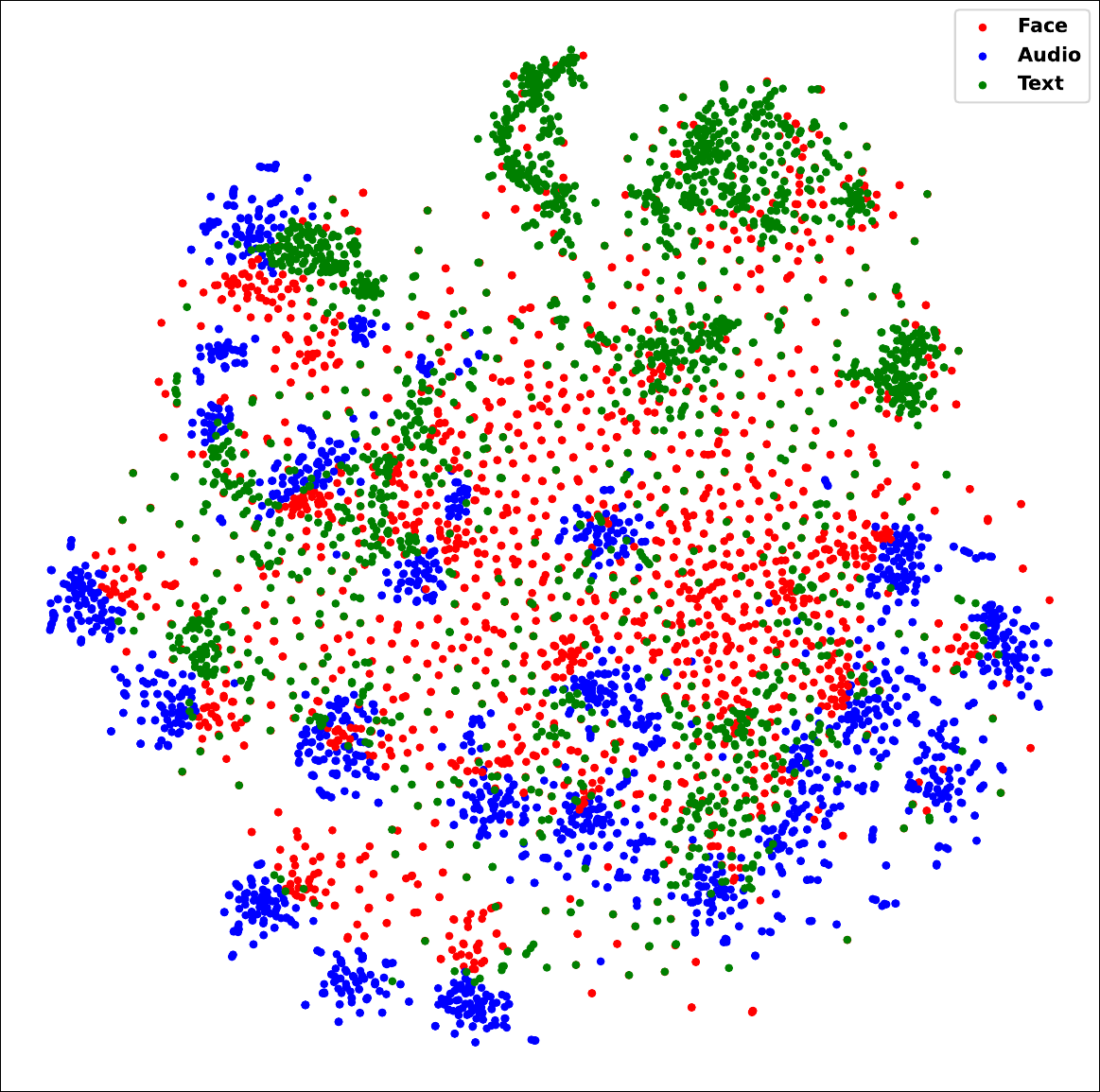} 
        \subcaption{MPE in flow matching}
    \end{minipage}
    \caption{TSNE visualization of MPE output embedding clustering.}
    \label{fig:append-mpetsne}
\end{figure}

As illustrated in Fig.~\ref{fig:append-mpetsne}, both MPEs exhibited similar trends: embeddings mapped by the MPE from different modalities reside within the same embedding space and are not partitioned into multiple subspaces where partitioning into subspaces would imply that each modality is encoded separately, failing to capture the intermodal relationships. Furthermore, the embeddings from audio prompts demonstrated stronger clustering, indicating that audio prompts are more directional than text and facial prompts. In contrast, text and facial prompts exhibit a one-to-many relationship with voice attributes, showing more significant variability.

\end{document}